\documentstyle[prb,preprint,aps]{revtex}
\begin{document}
\draft
\title{  Theory for the  Doping Dependence\\
         of  Spin Fluctuation Induced Shadow States\\
       in High-T$_{c}$ Superconductors}

\author{M. Langer, J. Schmalian, S. Grabowski, and K.H. Bennemann}
\address{Institut f\"ur Theoretische Physik,
  Freie Universit\"at Berlin, Arnimallee 14, \\
       14195 Berlin, Germany}
\date{October 4, 1995}
\maketitle
\begin{abstract}
We analyze the  doping dependence of the intensity
 and  energetical position of  shadow states
 in high -T$_{c}$  superconductors within
the 2D Hubbard model and
using our recently developed numerical method for the
 self consistent summation of bubble and ladder diagrams.
It is shown that  shadow states resulting from
  short range antiferromagnetic correlations occur
for small but finite excitation energies which decrease
for decreasing doping, reflecting a dynamically broken
symmetry with increasing lifetime.
Simultanously, the intensity of these new states increases,
the quasiparticle dispersion is strongly flattened,
and a pseudogap in the density of states occurs.
Finally, we discuss the importance of flat bands at
the Fermi level and nesting of the Fermi surface as general
prerequisites for the observability of shadow states.
  \end{abstract}

\newpage

The electronic excitation spectrum of high-T$_c$
superconductors and its relation to
short range antiferromagnetic correlations are currently of special
interest, since it reveals important information
about  the possibility of a spin fluctuation induced
pairing mechanism.
Here, the observation
of shadows of the Fermi surface (FS) for
$Bi_2 Sr_2 Ca Cu_2 O_{8+\delta}$ ~\cite{AOS94}  by Aebi {\em et al.}
is  intensively debated~\cite{C95,A95,KS90,LSG95,HMD95}.
Using  a new experimental technique to perform a FS mapping,
 Aebi {\em et al.} observed besides the main FS a shadow  with lower intensity
which is shifted by the wave vector ${\bf Q}=(\pi,\pi)$ with respect to
the main FS.
Due to the pronounced antiferromagnetic correlations
of the cuprates~\cite{ISY93,BGJ88} and the absence of any indication
of a structural origin of these new states in LEED experiments~\cite{AOS94},
the corresponding $(2 \times 2)$ superstructure is believed
to be of antiferromagnetic
origin~\cite{AOS94,KS90,LSG95,HMD95}.
Nevertheless, it is {\em a priori} not clear whether the antiferromagnetic
correlations  with
rather short correlation length are sufficient to produce new states, which
one would have expected only for systems with long range order~\cite{C95}.
Therefore,  important information about the magnetic origin of the shadow
states might   result from an experimental as well as theoretical
 investigation   of the doping dependence of the shadow
band intensity.

In this paper we present our results for the doping dependence of the
spectral density and consequently of the shadow states using
a new numerical method for the self consistent solution of the fluctuation
exchange approximation (FLEX)~\cite{BS89,BSW89} of the one-band Hubbard
model on the real frequency axis.
We show that the  intensity and the lowest
energy distance of the shadow states with respect to the Fermi level
 are  closely intertwined
and  depend sensitively on the doping concentration.
This is interpreted as a dynamical symmetry breaking where the
lifetime of the symmetry broken state   increases for decreasing
doping concentration.
Consequently, we argue that in the cuprates the transition from the
paramagnetic
to the antiferromagnetic state occurs on finite time scales rather
gradually,  although  a sharp phase transition  is still  observable
for even lower doping concentrations.

In order to achieve an appropriate description of the
low energy spin fluctuations we
use in the following the fluctuation exchange approximation (FLEX)
introduced by Bickers and Scalapino~\cite{BS89,BSW89}.
This self consistent summation of all bubble and ladder diagrams
is a conserving approximation in the sense of Baym and Kadanoff~\cite{BK61}.
In particular, it takes into account the interaction of the quasi particles
with spin fluctuations which are expected to be the dominating low
energy excitations in the high-$T_c$ materials.

We consider the one-band Hubbard Hamiltonian
\begin{equation}
H=\sum_{ij \sigma} (t_{ij}-\mu \delta_{ij})
   c^\dagger_{i \sigma}c_{j \sigma}
   +U \sum_i c^\dagger_{i \uparrow }c_{i \uparrow }
    c^\dagger_{i \downarrow }c_{i \downarrow } \, ,
\label{Hub}
\end{equation}
where, $c^\dagger_{i \sigma}$ is the creation operator of an electron
at lattice site $i$ and with spin $\sigma$.
$t_{ij}$ is the hopping matrix element between sites $i$ and $j$,
$\mu$ is the chemical potential and $U$ is the local
Coulomb repulsion.
The results presented below are obtained for nearest neighbor hopping
$t=0.25 \, {\rm eV}$, $U=4t$ and temperature $T=63\, {\rm K}$.
The self-energy of this Hamiltonian within the FLEX-approximation,
neglecting the particle-particle excitations which were shown to be of minor
importance~\cite{BSW89},
is given by the following momentum
and Matsubara-frequency  sum~\cite{BS89}
\begin{equation}
\Sigma_{{\bf  k} }(i\omega_n)=\frac{T}{  N} \sum_{{\bf  k}',n'}
 G_{{\bf k}'}(i\omega_{n'})
V_{{\bf k}-{\bf k}'}(i\omega_n-i\omega_{n'}) \, ,
\label{selfe}
\end{equation}
where,  $N$ is the number of lattice sites.
The effective interaction $V_{{\bf k}}(i\nu_m)$ resulting from the
summation of bubble and electron-hole ladder diagrams is given by
\begin{eqnarray}
V_{{\bf q} }(i\nu_m )=  \frac{U^2}{2} \chi^o_{{\bf q} }(i\nu_m )
\left(\frac{3}{1-U\chi^o_{{\bf q} }(i\nu_m )}
  +    \frac{1}{1+U\chi^o_{{\bf q} }(i\nu_m )} - 2 \right) \, .
\label{effint}
\end{eqnarray}
Here, $\chi^o_{{\bf q} }(i\nu_m )= -\frac{T}{  N} \sum_{{\bf  k},n}
 G_{{\bf k}+{\bf q}}(i\omega_{n}+i\nu_m)
  G_{{\bf k} }(i\omega_{n} ) $ is the particle-hole bubble and
 $ \omega_n=(2n+1)\pi  T $ and  $ \nu_m=2m\pi  T $ are the fermionic
and bosonic Matsubara frequencies, respectively.
Furthermore, the Greens function is given by the Dyson equation
$G_{{\bf  k} }(i\omega_n) = \left( i\omega_n + \mu - \varepsilon({\bf k})
                         - \Sigma_{{\bf  k} }(i\omega_n) \right) ^{-1}$
where $\varepsilon({\bf k}) = -2t (\cos(k_{x}) + \cos(k_{y}))$
is the free dispersion.
These equations are analytically transformed to the real frequency axis
yielding a set of equations for the Greens function
$G_{{\bf  k} }(\omega)$ and the self-energy
$\Sigma_{{\bf  k} }(\omega)$, which is solved
self consistently~\cite{SLG95}.

In Fig.~\ref{fig1}(a) we show our results for the spectral density
$\varrho_{\bf k}(\omega)=-\frac{1}{\pi} {\rm Im}G_{{\bf  k} }(\omega)$
for various ${\bf k}$-values shifted by ${\bf Q}$ with respect  to the main FS
as can be seen in the  inset, where we plot the first quadrant
of the Brillouine zone. The doping concentration  $x=1-n$, with
occupation  number per site $n$, is given by $x=0.12$.
Besides the main peak with a  position  similar to that of an uncorrelated
system, we find new states below the Fermi surface.
These are the shadow states which occur for ${\bf k}$-values where one
expects the states of the new band  of a long range ordered antiferromagnet,
although our calculation was performed in the paramagnetic state.
Furthermore, an estimate of the antiferromagnetic correlation length from
the spin spin susceptibility, neglecting vertex corrections,
yields $\xi \approx 2.5$ lattice spacings,
in agreement with neutron scattering experiments~\cite{TGS92}.
These results show that indeed spin fluctuation induced shadow states
without long range antiferromagnetic order and with short correlation length
exist.~\cite{LSG95}
The  distance  of these states with respect to the Fermi energy is always
finite,
i.e. they   never cross the Fermi energy, but are of the order of the
experimental
resolution
of a photoemission experiment.

In Fig.~\ref{fig1}(b) we present our results for the doping dependence
of  $\varrho_{\bf k}(\omega)$ at the FS-shadow near the ${\bf k}$-value
$(\pi,0)$.
One clearly recognizes the strong increase of the shadow band intensity
for decreasing doping concentration.
Consequently, an anomalous occupation of states with momentum
outside the FS occurs.
Besides the intensity variation, important information about the dynamical
character of the antiferromagnetic correlations can be obtained from
the position of these new states.
Therefore, we consider the quasi particle dispersion obtained from the
momentum variation of all local maxima of $\varrho_{\bf k}(\omega)$.

In Fig.~\ref{fig2} we show  our results for the
quasi particle dispersion (solid line and squares) for different doping
values in comparison with the dispersion of the corresponding
uncorrelated systems (dotted line) for ${\bf k}$-values indicated in the
inset.
The solid line refers to the global maximum and shows the main band dispersion,
 whereas the shadow states are indicated by the squares.
Most interestingly, one observs for decreasing doping
a reduction of the lowest  distance $\omega_{\rm exc}$
of a shadow state  with respect to the Fermi energy.
For $\omega_{\rm exc} \rightarrow 0$, we expect a static antiferromagnetic
order.
Therefore, we interpret the inverse of $\omega_{\rm exc}$ as a measure of the
life
 time of the dynamically symmetry broken state.
Another precursor effect of the antiferromagnetic state in the paramagnetic
phase
is the opening of a spin density gap, as can be seen from the strong
deformation
of the main band (solid line) upon doping.
For larger doping, solely a flattening of the dispersion near the Fermi level
occurs,
whereas a sudden jump   near the magnetic Brillouine zone border
$(\pi,0)$  can be seen for $x=0.12$ and $x=0.09$.
Consequently, on  a finite time scale, a quasi particle cannot distinguish the
actual
state from that of a long range ordered phase.
Therefore, the system behaves on shorter
time scales as in a symmetry broken state, although no antiferromagnetic
order occurs in the thermodynamic limit,
where each excitation has infinite time to relax into equilibrium.
This point of view is supported by the experimental observation by
Osterwalder {\em et al.}~\cite{OAS94} where the shadow states
seem to be more sharp for an    energy
$\omega = 60 \, {\rm meV}$   than  directly at the Fermi level.

In Fig.~\ref{fig3}, we present our results for the doping dependence of the
"van Hove" scale $\omega_{\rm vH}$, i.e. the main band   energy for
${\bf k}=(\pi,0)$, $\omega_{\rm exc}$ and the relative intensity of the shadow
states
with respect to the main band height.
Comparing $\omega_{\rm vH}$ with the corresponding energy for $U=0$,
we find  that due to the electronic correlations states are shifted towards
the Fermi level, and our results for $x \approx 0.16$ become comparable
with the  experimental  value of $ 30 - 50 \, {\rm meV}$~\cite{SD95}.
For lower doping values, a saturation of $\omega_{\rm vH}$, related to the
opening of a spin density gap, can be observed.
Furthermore, the close relation between  the variations of the  intensity
of the shadow states and  of $\omega_{\rm exc}$
can be seen in Fig.~\ref{fig3}(b) and (c).
Between $x=0.14$  and $x=0.12$ the abrupt decrease of
 $| \omega_{\rm exc}|$  occurs simultaneously with
the strong increase of the shadow band intensity.
Consequently, we find that the shadow band phenomenon appears suddenly
 below a critical doping concentration ($ x \approx 0.14$).
Note that this might  also  be related
to the optimal doping $x_{\rm opt}=0.13$ of the superconducting transition
temperature $T_c$, which was found by us in a corresponding FLEX-calculation
in the superconducting state~\cite{GSL95}.

Although all  our calculations are performed using a model
dispersion, we believe that the
occurrence of shadow states is of more general character.
The two important prerequisites for
these states are a flat dispersion of the quasiparticles near the
Fermi energy and flat portions
of the Fermi surface, i.e. FS-nesting.
These features are experimentally observed in various cuprate
systems~\cite{SD95} and are  also a
basic content of more realistic parametrizations of band structure
calculations~\cite{ALJ95}.
While FS-nesting favors the antiferromagnetic correlations,
flat bands are necessary
because the states with momentum ${\bf k}$ and  ${\bf k} + \xi^{-1}$, which
contribute both
to a shadow state at ${\bf k}+{\bf Q}$, have to be energetically
close enough together to ensure a well defined peak in
$\varrho_{{\bf k}+{\bf Q}}(\omega)$.
Even if a  change of the model dispersion and the FS leads to
less pronounced nesting,
a correlation induced flattening of the bands can compensate this
with respect to the shadow
state intensity.
It is important to remark that our theory yields for larger values
of the ratio $U/t$  an increased  range   of flat bands in ${\bf k}$-space.
For $U/t=6$ and $x=0.12$ the quasi particle   energy for
 ${\bf k}=(5 \pi/8,0)$ is $80 \, {\rm meV}$ in comparison with $130 \, {\rm
meV}$ for
$U/t=4$.
Consequently, we believe that the moderate values of $U/t$ of our calculation
give a lower bound for  the shadow band  intensity for the model dispersion
under consideration.

In conclusion,  the doping dependence of spin fluctuation induced shadow states
 was investigated  using a self consistent description for the
dynamical excitation spectrum
of the 2D one  band Hubbard model within the FLEX-approximation.
It was shown that below a critical doping concentration $x \approx 0.14$ these
states
appear suddenly simultaneously with a sharp decrease of their lowest
energy distance $\omega_{\rm exc}$ with respect to the Fermi energy.
The additional opening of a spin density gap makes it impossible to
discriminate
on finite time scales (shorter than $\omega_{\rm exc}$) this paramagnetic state
with
strong dynamical spin fluctuations from a long range ordered state.
All this demonstrates the strong influence of short
range antiferromagnetic correlations on the excitation
spectrum of these compounds and might be stimulating for further
experimental investigations of the shadow band phenomenon.

%
%
%
%

\newpage

\begin{figure}
\caption{Spectral density of the shadow states for (a) doping
concentration $x=0.12$ and various ${\bf k}$-values shifted by ${\bf Q}$
with respect to the main Fermi surface and (b) for  ${\bf k}$
on the FS-shadow closest to $(\pi,0)$ and for different doping values.
The inset indicates the main Fermi surface (solid line) its shadow (dashed
line) and the ${\bf k}$-values of part (a)   in the first quadrant
of the Brillouine zone.}
\label{fig1}
\end{figure}
\begin{figure}
\caption{Quasi partice dispersion  of the main band (solid line) and the shadow
states (squares) for different doping values and for  ${\bf k}$-values
indicated
in the inset. The dashed line refers to the dispersion for $U=0$.
Note the pronounced deformation of the main band dispersion and the decrease
of the lowest  energy distance of a shadow state with respect to the Fermi
energy
($\omega=0$) for decreasing doping.}
\label{fig2}
\end{figure}

\begin{figure}
\caption{Doping dependence
(a) of    the main band energy $\omega_{\rm vH}$ for  ${\bf k}=(\pi,0)$
       in comparison with  its  value for $U=0$,
(b) of the lowest   energy distance $\omega_{\rm exc}$ of a shadow
      state with respect to the Fermi energy,   and
(c) of the relative  intensity of the  shadow state with respect to the main
      band height
      closest to $(\pi,0)$.
 Note the simultaneous and sudden appearance of large shadow
band intensities and low $\omega_{\rm exc}$ for doping values below $x=0.14$.}
\label{fig3}
\end{figure}

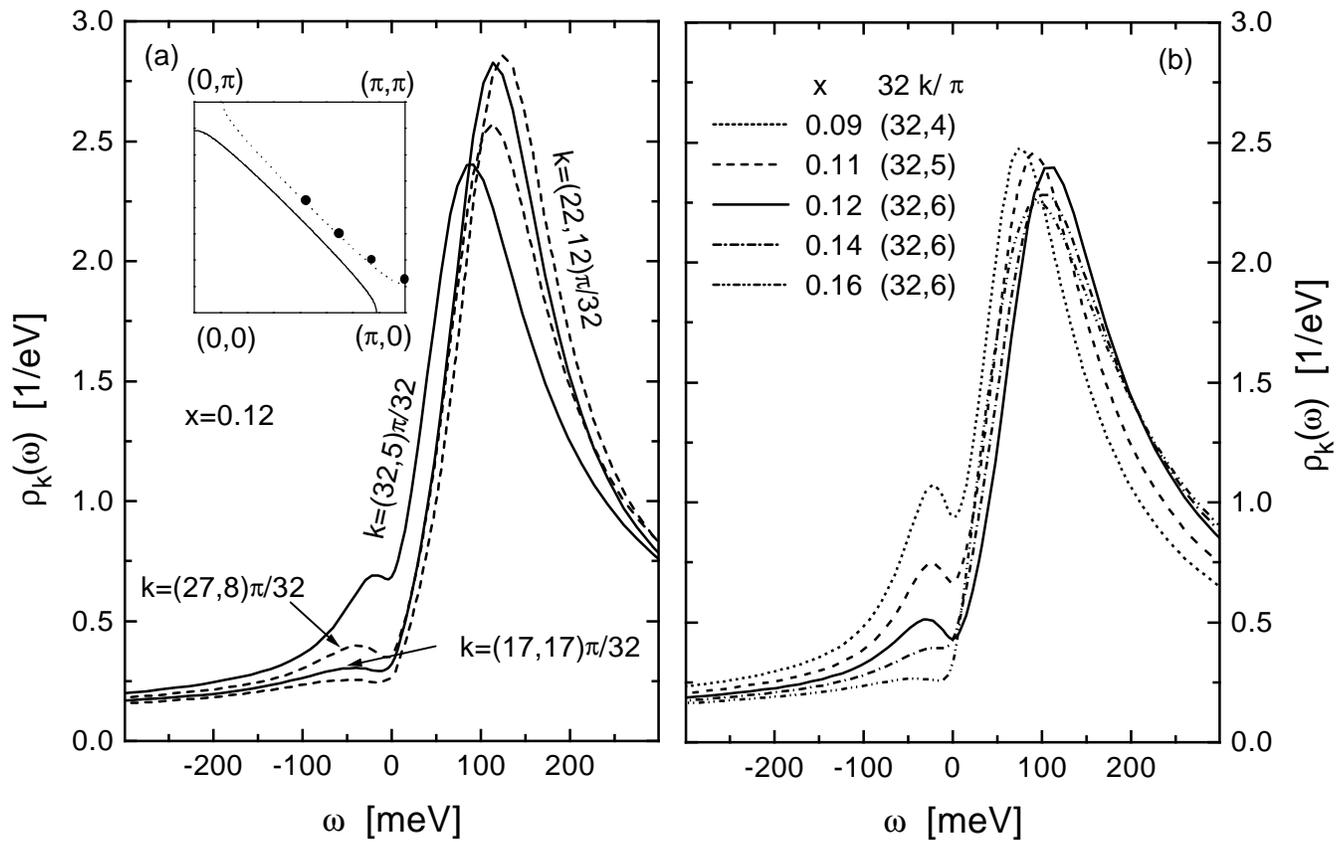
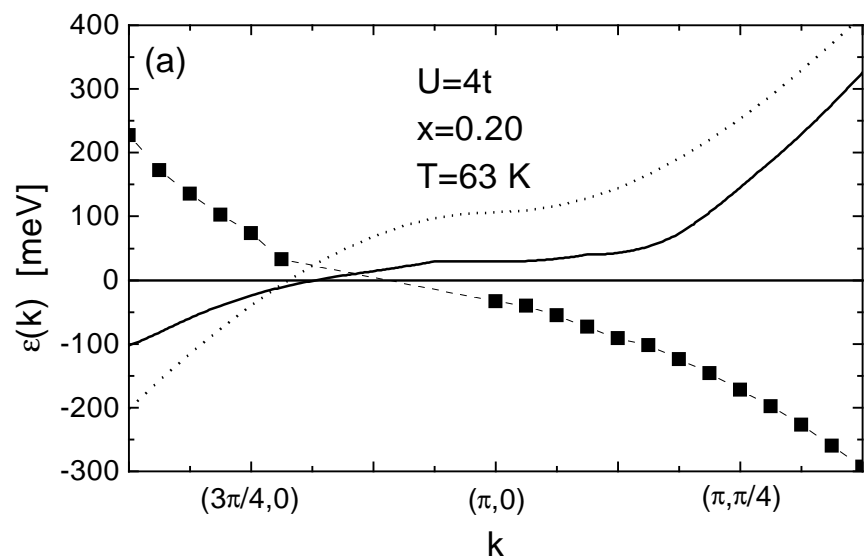
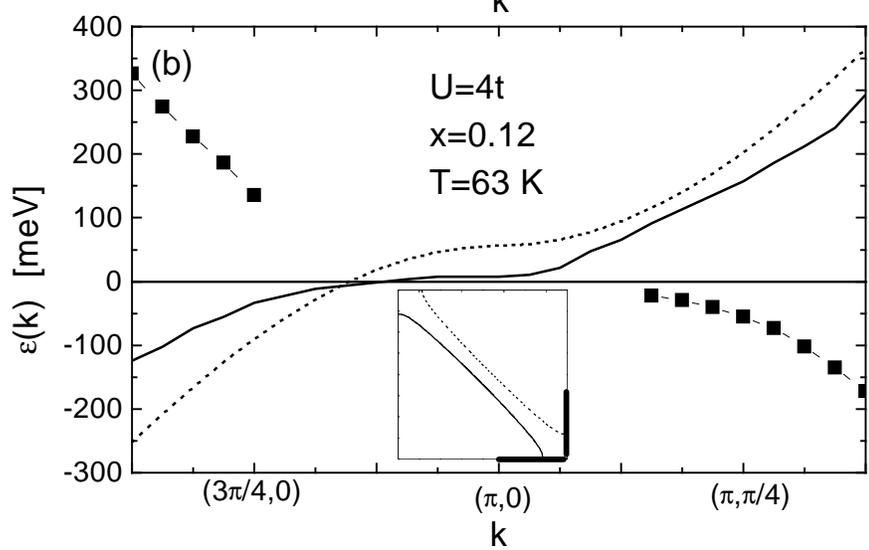
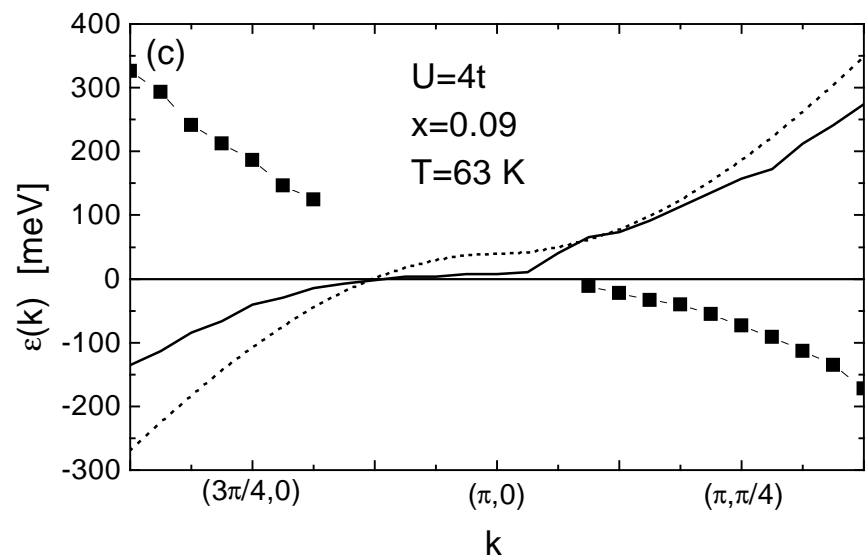
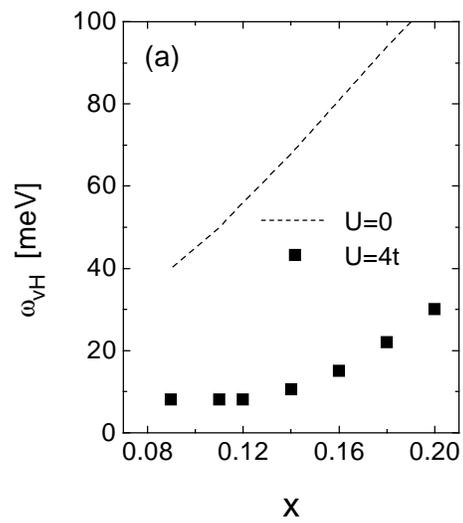
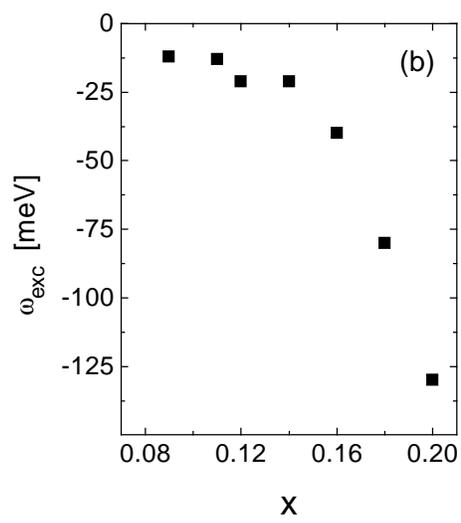
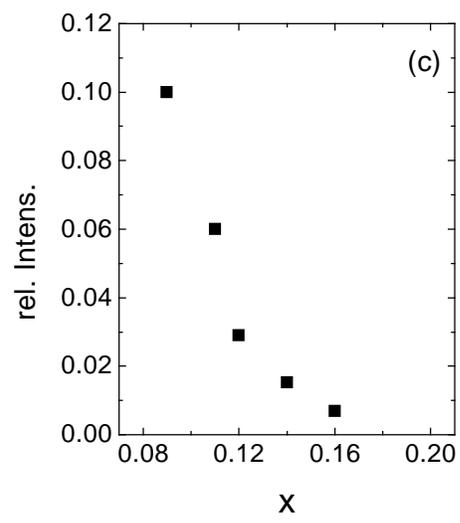
\end{document}